\documentclass{PoS}
\usepackage{wrapfig}
\title{Multi-nucleon bound states in $N_f=2+1$ lattice QCD}

\ShortTitle{Multi-nucleon bound states in $N_f=2+1$ lattice QCD}

\author{
   Takeshi Yamazaki$^{a}$,  
   Ken-Ichi Ishikawa$^b$,
   Yoshinobu Kuramashi$^{c,d,e}$,
   \speaker{Akira Ukawa$^{c,d}$}\thanks{E-mail:ukawa@ccs.tsukuba.ac.jp}
   \\
   \\
   \\
   \llap{$^a$}
Kobayashi-Maskawa Institute for the Origin of Particles and the Universe, Nagoya University, Naogya, Aichi 464-8602, Japan   
   \\
   \llap{$^b$}
Department of Physics, Hiroshima University, Higashi-Hiroshima, Hiroshima 739-8526, Japan.
   \\ 
   \llap{$^c$}
Graduate School of Pure and Applied Sciences, University of Tsukuba, Tsukuba, Ibaraki 305-8571, Japan
   \\
   \llap{$^d$}
Center for Computational Sciences, University of Tsukuba, Tsukuba, Ibaraki 305-8577, Japan
   \\
  \llap{$^e$}
RIKEN Advanced Institute for Computational Science, Kobe, Hyogo 650-0047, Japan
}

\abstract{We report on our on-going effort to calculate the properties of light nuclei directly from quarks and gluons based on lattice QCD.   After briefly introducing our motivations and aims, we describe our strategy of fixing the strange quark mass at its physical value and approaching the physical point for the up and down quark masses step by step from the region of heavy quark masses.  A successful calculation for the pion mass of 0.51~GeV is reviewed, and the status for a lighter pion mass of 0.30~GeV is reported. }

\FullConference{31st International Symposium on Lattice Field Theory - LATTICE 2013\\
		July 29 - August 3, 2013\\
		Mainz, Germany}

\begin{document}

\section{Motivations and Aims}

Since its inception more than a half century ago, the phenomenological description of atomic nuclei in terms of protons and neutrons and their interactions has been enormously successful.  
Because protons and neutrons are made of quarks and gluons, however, the very success of such an effective description has to be explained based on the underlying dynamics of QCD.  

This is not merely an exercise confirming the known facts in nuclear physics, but a prerequisite for exploring the more interesting region of nuclear physics.  Whenever one tries to explore unnatural nuclei such as those having a large neutron proton ratio or with quantum numbers other than isospin, phenomenological nucleon potentials,  which can be confirmed only for natural nuclei, always suffer from uncertainties.  Here lattice QCD is the only method which can be reliably employed to extract predictions on nuclei in their whole spectrum of quantum numbers. 

Another motivation for QCD-based nuclear physics stems from the fact that empirical nuclear models are simply not capable of predicting what would have happened if natural constants such as the quark masses were different from what they are in our universe.   

Those motivations were already in the back of mind in the early lattice QCD studies of nucleon-nucleon interactions almost two decades ago~\cite{fukugita1994}.  The calculations possible at the time, however, were quite limited, employing the quenched approximation of ignoring quark vacuum polarizations and with unnaturally heavy quark masses.  

Lattice QCD has made much progress since then, and the aim of our recent work~\cite{yamazaki2010,yamazaki2011,yamazaki2012}  is to systematically calculate the properties of atomic nuclei including the full effects of quark vacuum polarizations for the physical quark masses and on a lattice of a large enough physical size. 
In the present talk we present the status report of our endeavor.  See ref.~\cite{others} for some representative studies by other groups. 

\section{Issues}

The primary quantity we evaluate for atomic nuclei is their ground state energy or the mass for zero spatial momentum.   There are several issues in their evaluation to which we need to pay attention.  One of the issues, now well understood, is that a negative value of the energy difference $\Delta E_L=E_L-Am_N$, where $A$ is the mass number of the nucleus under investigation, $E_L$ its energy calculated on a lattice of a size $L$, and $m_N$ the nucleon mass, does not necessary mean a bound state formation.    We need to calculate ${\Delta} E_L$ for a set of lattice sizes $L$ and make sure that the limit ${\Delta} E_\infty=\lim_{L\to\infty}{\Delta} E_L$ converges to a non-zero negative value ${\Delta} E_\infty<0$.  The behavior ${\Delta} E_L\propto 1/L^3\to 0$ as $L\to\infty$, on the other hand, signifies that the state is a scattering state. 

\begin{figure}[!t]
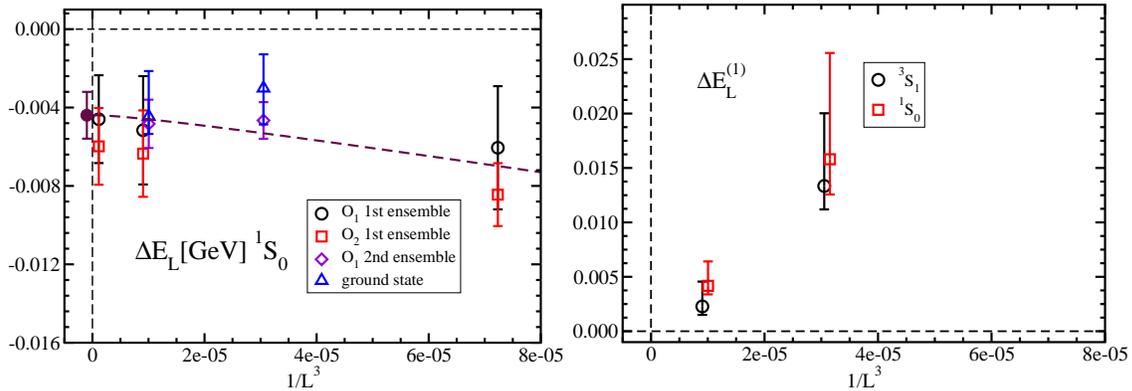

\includegraphics*[angle=0,width=0.49\textwidth]{Fig/DdE_10.eps}
\includegraphics*[angle=0,width=0.49\textwidth]{Fig/DdE_n1.eps}
\caption{
Left panel shows the volume dependence of the energy of the ground state of two nucleons in the spin singlet channel relative to that of two free nucleons.  Right panel shows that for the first excited state in the spin singlet and triplet (deuteron) channels.  Calculations are made in quenched QCD with a heavy pion mass of $m_\pi=0.81$~GeV. ~\cite{yamazaki2011}. 
\label{fig:two-nucleon}
}
\end{figure}

If the ground state is a bound state, a simple quantum mechanical argument shows that the first excited state is a scatting state just above the threshold.  Hence, if we calculate the energy of the first excited state $E_L^{(1)}$ by forming an orthogonal state to the ground state, then one expects to find a size dependence $\Delta E_L^{(1)}\approx \alpha/L^3\to 0$   {\it with a positive constant } $\alpha$.  For the two nucleon state, we have recently confirmed this behavior as shown in Fig.~\ref{fig:two-nucleon}, albeit for quenched QCD with a heavy pion mass~\cite{yamazaki2011}.   We also emphasize that the scattering length comes out negative, consistent with a bound state formation, only if one applies the Luescher's formula to the first scattering state~\cite{yamazaki2011}.   This point was later confirmed by NPLQCD Collaboration for $N_f=3$ QCD with a pion mass of $m_\pi\approx 0.8$~GeV~\cite{NPLQCD2013}. 

Another issue concerns the dependence on the quark mass.  For heavy quark masses,  majority of lattice QCD calculations to date, which directly evaluates the energy of multi-nucleon states from their correlation functions, yields negative values for the energy difference irrespective of the nucleon number and spin, indicating bound states.  For some channels, such as the spin singlet channel of the two-nucleon state, for which bound states do not exist for the physical quark mass, this poses an interesting dynamical issue that at some quark mass toward the physical value, those bound states have to disappear.  Since the physical quark masses are small but non-zero even for the light up and down quarks, one can stretch the issue and ask if nuclei are bound if all quarks have vanishing masses. 

The third issue, which is technical in nature,  is a factorial growth of the number of quark contractions in the evaluation of correlations functions for nuclei.  Up to helium with the mass number $A=4$, this problem has been handled by reducing the number of terms by making use of various symmetries and other techniques, {\it e.g.}, from the nominal $6!\times 6!=5123,000$ to $1,107$ terms for Helium 4~\cite{yamazaki2010}.  For heavier nuclei starting with $A=5$, more systematic methods are needed, and effort in this direction can be seen in recent literatures.~\cite{doi2013,detmold2013,gunther2013}     
 
\section{Strategy}

We employ the non-perturbatively $O(a)$ improved Wilson quark action with the clover term and the Iwasaki gluon action.  Extensive calculations have been carried out for this action combination~\cite{pacscs,phys_point},  from which precision information is available connecting the lattice parameters with quark masses and lattice spacing.  The current series of calculations are being carried out for the $N_f=2+1$ case at $\beta=1.90$ with $c_{SW}=1.715$.  The physical point inputs are $a^{-1}=2.194$~GeV, $am_s = 0.02934$  at $\kappa_{ud} = 0.13779625$ and $\kappa_s = 0.13663375$~\cite{phys_point}. 

We fix the strange quark mass defined in terms of the axial vector current at the physical value, and decrease the degenerate up and down quark masses step by step from heavy values toward the physical one.  In practice we have employed $m_\pi=0.51$~GeV, $0.30$~GeV so far.  At each pion mass, we repeat measurements for a set of lattice sizes to distinguish bound states from scattering states. 

The quark source is smeared by the function $\psi(r)=A\exp(-Br)$.  We adjust the smearing parameter  $B$ so that an early and good plateau is realized for the nucleon state.  This is important to ensure that we truly measure the energy difference between the interacting multi-nucleon state and the corresponding free state. 

To fully utilize the dynamical content of the gluon configuration across the spatial and temporal extent of the lattice, we make multiple measurements for each gluon configuration shifting the source location for both time and space directions.  In addition, for large lattice sizes such as $L=48$ and $ 64$, we generate configurations on a space-time symmetric lattice of a size $L^4$, and {\it make measurements in all four directions}.   The boundary condition is taken periodic in both space and time.  

\section{Results}

\begin{table}
\caption{Run statistics for (a) $m_\pi=0.51$~GeV~\cite{yamazaki2012} and for (b) $0.30$~GeV.  The latter runs are still in progress.
\label{tab:runs}
}
\begin{tabular}{lllllll}
\hline
(a) $m_\pi=0.51$~GeV\\
\hline
size&\# configs &traj sep&bin size&\# meas/conf&$m_\pi$(GeV)&$m_N$(GeV)\\
\hline
$32^3\times 48$& 200 & 20 & 10 & 192 &  0.5109(16) & 1.318(4) \\
$40^3\times 48$& 200 & 10 & 10 & 192 &  0.5095(8) & 1.314(4) \\
$48^3\times 48$& 200 & 10 & 20 & 192 &  0.5117(9) & 1.320(3) \\
$64^3\times 64$& 190 & 10 & 19 & 256 &  0.5119(4) & 1.318(2) \\
\hline
(b) $m_\pi=0.30$~GeV\\
\hline
$48^3\times 48$& 360 & 10 & 20 & 576 &  0.3004(15) & 1.058(2) \\
$64^3\times 64$& 160 & 10 & 10 & 384 &  0.2985(8) & 1.057(2) \\\hline
\end{tabular}
\end{table}

\begin{figure}[!t]
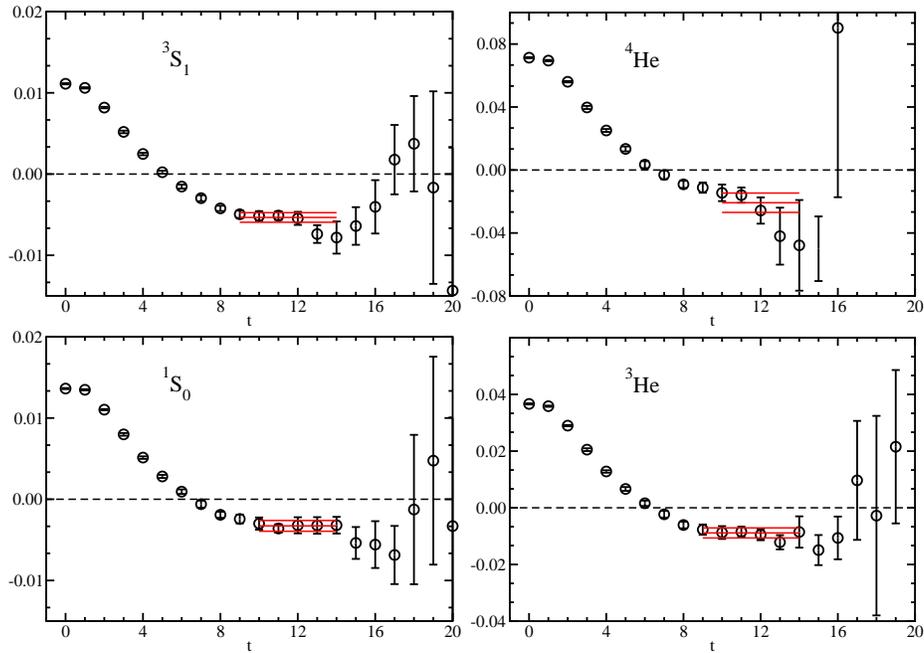

\begin{center}
\includegraphics*[angle=0,width=0.4\textwidth]{Fig/eff_R_3S1.eps}
\includegraphics*[angle=0,width=0.4\textwidth]{Fig/eff_R_4He.eps}\\
\includegraphics*[angle=0,width=0.4\textwidth]{Fig/eff_R_1S0.eps}
\includegraphics*[angle=0,width=0.4\textwidth]{Fig/eff_R_3He.eps}
\caption{
Effective energy difference for the size $L=64$ for $m_\pi=0.51$~GeV.  Left panels shows results for the two nucleon system, and right panels those for Helium 4 and 3. 
\label{fig:deltaenergympi051L64}
}
\end{center}
\end{figure}

\begin{figure}[!t]
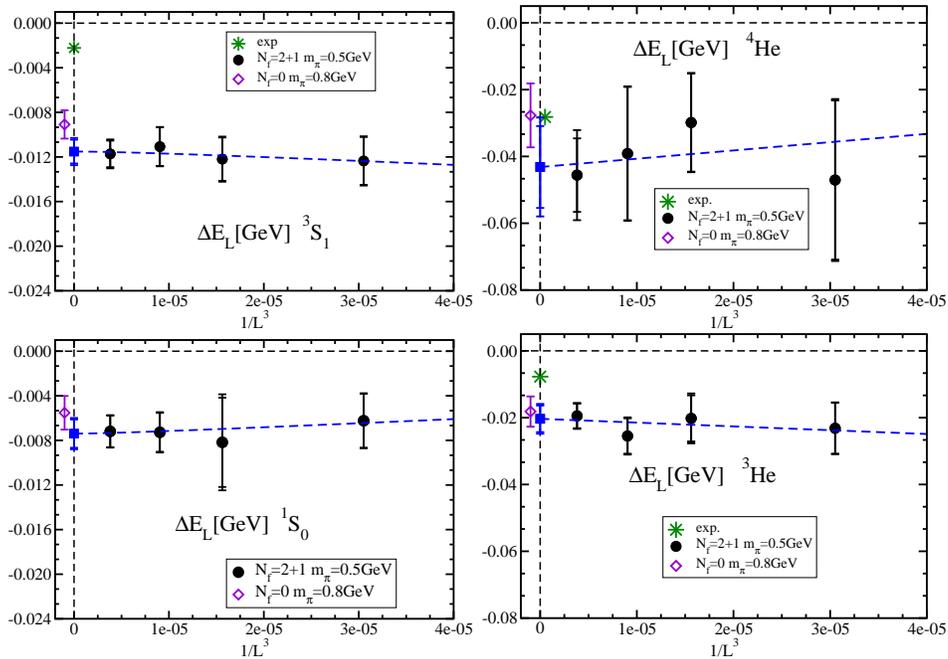

\begin{center}
\includegraphics*[angle=0,width=0.41\textwidth]{Fig/dE_31.eps}
\includegraphics*[angle=0,width=0.41\textwidth]{Fig/dE_h4.eps}\\
\includegraphics*[angle=0,width=0.41\textwidth]{Fig/dE_10.eps}
\includegraphics*[angle=0,width=0.41\textwidth]{Fig/dE_h3.eps}
\caption{
Volume dependence of the energy difference for two nucleon states for $m_\pi=0.51$~GeV (left panels) and for Helium 4 and 3 (right panels).  
\label{fig:bind051}
}
\end{center}
\end{figure}

The run statistics is listed in Table~\ref{tab:runs}. 
Let us start from results for a heavier pion mass of $m_\pi=0.51$~GeV.  
We define an effective energy difference as 
\begin{equation}
\Delta E(t) = -\log \frac{\delta G(t)}{\delta G(t-1)}\quad {\rm with} \quad \delta G(t) =  \frac{G_A(t)}{G_N(t)^A},
\end{equation}
where $G_A(t)$ and $G_N(t)$ denote the nuclear correlation function with a mass number $A$ and the nucleon correlation function, both projected for zero spatial momentum.  The effective energy difference for various channels is illustrated in Fig.~\ref{fig:deltaenergympi051L64} for the size $L=64$. 
We observe reasonable plateaux in all channels around $t\approx 9-14$, and we make fits over the temporal intervals indicated by the band of three lines.  The volume dependence of the energy difference is shown in Fig.~\ref{fig:bind051}.   The energy difference is almost constant against spatial volume for all channels shown, leading to  negative values if extrapolated linearly in $1/L^3$ to infinite volume.  Therefore all channels examined in Figs.~\ref{fig:bind051} are bound states at $m_\pi=0.51$~GeV.

\begin{figure}[!t]
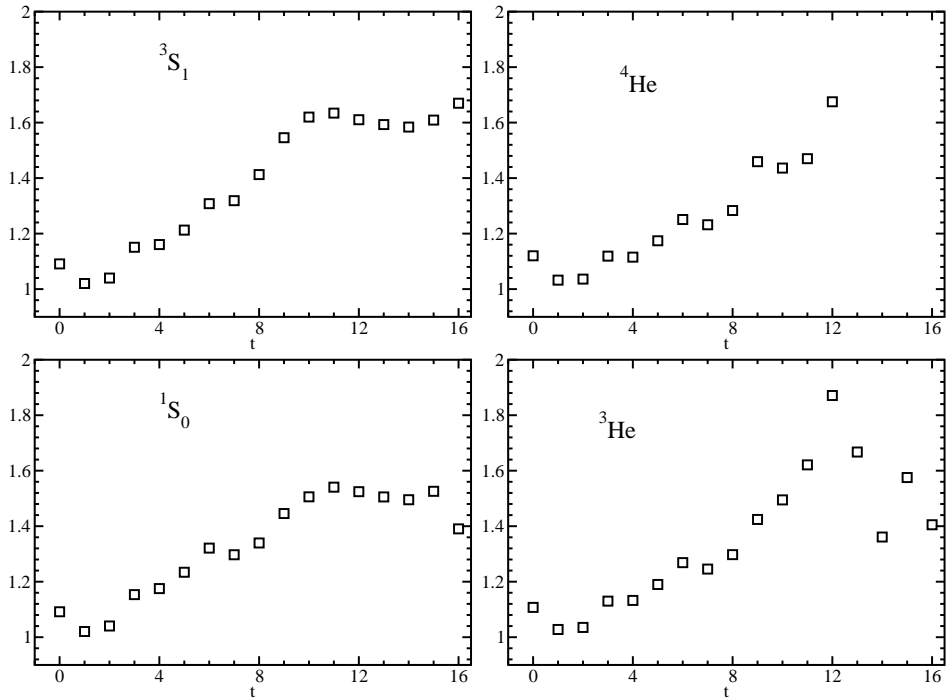

\begin{center}
\includegraphics*[angle=0,width=0.41\textwidth]{Fig/re_3S1.eps}
\includegraphics*[angle=0,width=0.41\textwidth]{Fig/re_4He.eps}\\
\includegraphics*[angle=0,width=0.41\textwidth]{Fig/re_1S0.eps}
\includegraphics*[angle=0,width=0.41\textwidth]{Fig/re_3He.eps}
\caption{
Ratio of relative error of nuclear correlation function
$r(t)$ calculated for $m_\pi=0.30$~GeV and 0.51~GeV for the size $L=48$ for the statistics listed in Table 1.  Left panels are for the two nucleon states, and right panels for Helium 4 and 3.
\label{fig:errorratio}
}
\end{center}
\end{figure}

Currently we are running calculations for the pion mass of $m_\pi=0.30$~GeV.  
As might have been expected, we are finding that increased statistics is required.  In order to quantify the situation, we calculate the relative error of nuclear correlation functions ${\delta_A}(t) = \delta G_A(t)/G_A(t)$ at $m_\pi=0.30$~GeV and 0.51~GeV for the lattice size $L=48$, and plot their ratio $r(t)=\delta_A(t)\vert_{m_\pi=0.30~{\rm GeV}}/\delta_A(t)\vert_{m_\pi=0.51~{\rm GeV}}$ in Fig.~\ref{fig:errorratio}.  At the temporal range of $t\approx 9-14$ where we expect to extract the energy difference, the ratio $r(t)$ takes values around 1.5.  Therefore, to bring the ratio down to unity, roughly twice the statistics compared to the present number of configurations will be required at $m_\pi=0.30$~GeV.  If we measure statistics in terms of the number of configurations times the number of measurements on each configuration, this means a factor of 12 increase for $m_\pi=0.30$~GeV as compared to that for $m_\pi=0.50$~GeV.  

\section{Summary}

We have reported on our ongoing effort at extracting light nuclear properties directly from lattice QCD calculations at the physical point.  Our calculations at $m_\pi=0.51$~GeV has been completed already~\cite{yamazaki2012}, and our runs are continuing for a lighter mass of $m_\pi=0.30$~GeV, for which we hope to report the results in the near future.  

\section*{Acknowledgements}
Numerical calculations for the present work have been carried out
on the HA8000 cluster system at Information Technology Center
of the University of Tokyo, on the PACS-CS computer 
under the ``Interdisciplinary Computational Science Program'' of 
Center for Computational Sciences in University of Tsukuba, 
on the T2K-Tsukuba cluster system and HA-PACS system at University of Tsukuba,
and on K computer at RIKEN Advanced Institute for Computational Science.
We thank our colleagues in the PACS-CS Collaboration for helpful
discussions and providing us the code used in this work.
This work is supported in part by Grants-in-Aid for Scientific Research
from the Ministry of Education, Culture, Sports, Science and Technology 
(Nos. 18104005, 18540250, 22244018, 25800138) and 
Grants-in-Aid of the Japanese Ministry for Scientific Research on Innovative 
Areas (Nos. 20105002, 21105501, 23105708).

\end{document}